\documentclass[doublecol]{epl2} 
\usepackage{bm}
\usepackage{graphicx}
\usepackage{leftidx}
\usepackage{pict2e}

\title{Low Temperature Mass Spectrum in the Ising Spin Glass}

\author{A. Crisanti\inst{1}
  \and
        C. De Dominicis\inst{2}
}

\institute{
  \inst{1}Dipartimento di Fisica, Universit\`a di Roma 
              {\em La Sapienza} and  ISC-CNR, 
              P.le Aldo Moro 2, I-00185 Roma, Italy.\\
  \inst{2} {\em Institut de Physique Th\'eorique}, CEA -
              Saclay - Orme des Merisiers, 91191 Gif sur Yvette, 
              France
}

\abstract{
We study the spectrum of the Hessian of the Sherrington-Kirkpatrick model
near $T=0$, whose eigenvalues are the masses of the bare propagators in 
the expansion around the mean-field solution.
In the limit $T\ll 1$ two regions can be identified. The first for $x$ close
to $0$, where $x$ is the Parisi replica symmetry breaking scheme parameter.
In this region the spectrum of the Hessian is not trivial, and maintains the 
structure of the full replica symmetry breaking state found at higher 
temperatures.
In the second region $T\ll x \leq 1$ as $T\to 0$, the bands typical of the
full replica symmetry breaking state collapse and only two 
eigenvalues are found: a null one and a positive one.
We argue that this region has a droplet-like behavior.
In the limit $T\to 0$ the width of the full replica symmetry breaking 
region shrinks to zero and only the droplet-like scenario survives.
}

\date{V 1.4.6 2010/09/09 11:31:44 AC}

\begin{document}

\maketitle

The physics of spin glasses is still an active field of research because
the methods and techniques developed to analyze the static and dynamic 
properties have found applications in a variety of others fields of the
complex system world, such as
neural networks or combinatorial optimization or glass physics. 
In the study of spin glasses a central role is played by 
the Sherrington-Kirkpatrick (SK) model \cite{SheKir75},
introduced in the middle of 70's, as a mean field model for spin glasses.
Despite the fact that its solution, 
known as the ``Parisi solution'' \cite{Par79,Par80c},
was found $30$ years ago, some aspect are still far from
being completely understood. 
In this Note we discuss the stability,
that is the eigenvalue spectrum of the Hessian of the fluctuations,
of the Parisi solution in the $T\ll 1$ and its implications for the 
Replica Symmetry Breaking (RSB) versus droplet scenarios.

{\sl Model.}--The model is defined by the Hamiltonian \cite{EdwAnd75}
\begin{equation}
  H = -\frac{1}{2}\sum_{i,j} J_{ij}\,s_i\,s_j
\end{equation}
where $s_i=\pm 1$ are $N$ Ising spins located on a regular $d$-dimensional 
lattice and the symmetric bonds $J_{ij}$, which couple nearest-neighbor 
spins only, are random quenched Gaussian
variables of zero mean. 
The variance is properly normalized to ensure a well defined thermodynamic 
limit $N\to\infty$.
By using the standard replica method to average over the disorder, 
the free energy density $f$ in the thermodynamic limit 
can be written
as function of the symmetric $n\times n$ site dependent 
replica overlap matrix $Q_i^{ab}$ as
$-\beta f = \lim_{n\to 0} \frac{1}{n} \max_{Q} {\cal L}[Q]$
with \cite{BraMoo79} 
\begin{eqnarray}
{\cal L}[Q] &=& -\frac{\beta^2}{2}\sum_{\bm q}(q^2+1)
              \sum_{(ab)}\,(Q_{\bm q}^{ab})^2 
\nonumber\\
 &\phantom{=}&\phantom{==}      
    + \sum_i \ln{\rm Tr}\exp\Bigl({\beta^2\sum_{(ab)}Q_i^{ab}s^as^b}\Bigr)
\end{eqnarray}
where $Q_{\bm q}^{ab}$ is the spatial Fourier Transform of $Q_i^{ab}$ and
$\beta = 1/T$. The notation ``$(ab)$'' means that only distinct 
ordered pairs $a < b$ ($a,b = 1,\ldots,n$)
are counted.
By writing $Q_i^{ab} = Q^{ab} + \delta Q_{i}^{ab}$ and expanding 
${\cal L}[Q]$ in powers of $\delta Q_{i}^{ab}$ one generates the loop expansion.
The site-independent
$Q^{ab}$ is given by the {\sl mean field} value
$Q^{ab} = \langle s^as^b\rangle$, where 
angular brackets denote a weighted average with
$\exp (\beta^2\sum_{(ab)}Q^{ab}s^as^b)$.
This follows from the stationarity of ${\cal L}[Q]$, that is the vanishing 
of the linear term in the expansion, and 
ensures that no tadpoles are present.
The quadratic term of the expansion defines the bare propagators 
whose ``masses'' are
given by the eigenvalues of the non-kinetic part of the fluctuation
matrix:
\begin{equation}
\label{eq:hes}
M^{ab;cd} = \delta_{(ab);(cd)}^{Kr} - \beta^2\Bigl[
           \langle s^a s^b s^c s^d\rangle - 
           \langle s^a s^b\rangle\langle s^c s^d\rangle 
             \Bigl]
\end{equation}
that is the Hessian matrix of the SK model.

Stability of the Parisi solution for the SK model near its critical 
temperature $T_c$, has been established long ago by 
exhibiting the eigenvalues of the Hessian matrix \cite{deAlmTho78,DeDomKon83}.
In few words, one has a Replicon band whose 
lowest masses are zero modes, and a Longitudinal-Anomalous band, sitting 
at $(T_c - T)$, of positive masses, both with a band width of order 
$(T_c - T)^2$.
The analysis was extended via the derivation of Ward-Takahashi 
identities \cite{DeDomKonTem98}, showing that the zero Replicon modes 
would remain null in the whole low temperature phase, and hence
would not ruin the stability under loop corrections to the mean field solution.

Despite these efforts a complete analysis of the stability in the zero 
temperature limit is still missing. Near 
$T_c$ one can take advantage of the vanishing of the order parameter for
$T=T_c$ and expand ${\cal L}[Q]$, a simplification clearly missing 
close to zero temperature, where the order parameter stays finite.

{\sl Low Temperature Phase.}--As the temperature is lowered the ergodicity 
breaks down at the critical 
temperature $T_c=1$. Below $T_c$
the phase of the SK model is characterized by a large, yet not
extensive, number of degenerate locally stable states in which the system
freezes. The symmetry under replica exchange is broken
and the overlap $Q^{ab}$ becomes a non-trivial function of replica indexes.
Assuming $R$ steps of RSB the matrix $Q^{ab}$ is
divided, following the Parisi parameterization \cite{Par80c},
into successive boxes of decreasing size $p_r$, with $p_0=n$ and 
$p_{R+1}=1$, with elements given 
by\footnote{        For consistency one takes
        $Q^{aa} = Q_{R+1} = 1$.
}
\begin{equation}
 Q^{ab} = Q_{r}, \qquad r = 0,\ldots, R+1
\end{equation}
where $r = a\cap b$ denotes the overlap between replicas $a$ and $b$.
This means that $a$ and $b$ belongs to the same box of size 
$p_r$, but to two distinct boxes of size $p_{r+1}<p_r$.
The solution of the SK model is obtained by taking $R\to\infty$.
In absence of an external field the overlap $Q^{ab}$ takes values
between zero and a maximum value $q_c(T) \leq 1$, and so for $R\to\infty$ 
the matrix $Q^{ab}$ is described by a continuous non-decreasing function
$Q(x)$ parameterized by the variable $x$. In the Parisi 
scheme  $x\in[0,1]$ and gives the probability for a pair of 
states to have an
overlap $Q^{ab}$ not larger than $Q(x)$. 

The meaning of $x$ depends on the parameterization used for the
matrix $Q^{ab}$. 
In the dynamical approach \cite{Som81,CriHorSom93,CriLeu07} $x$ 
labels the relaxation 
time scale $t_x$, so that $Q(x) = \langle s(t_x)\,s(0)\rangle$.
Here the angular brackets denotes time (and disorder) averaging.  
The smaller $x$ the longer $t_x$.
All time scales diverges in the thermodynamic limit but 
$t_{x'}/t_x \to \infty$ if $x > x'$.
To make contact with the static Parisi solution one takes $x\in[0,1]$, 
with $x=0$ corresponding to the largest possible relaxation time
and $x=1^{-}$ to the shortest one. With this assumption one recovers
$Q(0)=0$ and $Q(1^-)= q_c(T)$. In both cases $Q(1)=1$, since it gives
the self or equal-time overlap. 
Other choices are possible, e.g., those used in Refs.
\cite{SomDup84,OppShe05,SchOpp08,OppSch08} to handle
the $T\to 0$ limit.
We stress however that different choices just give a 
different parameterization
of the function $Q(x)$, but do not change the physics, since the relevant 
quantities are the possible values $q$ that the function $Q(x)$ can take 
and their probability distribution $P(q)$. This property is
called {\sl gauge invariance} \cite{Som81,DeDomGabDup82,SomDup84}.
In what follows we assume the Parisi parameterization. 

\begin{figure}
\includegraphics[scale=0.9]{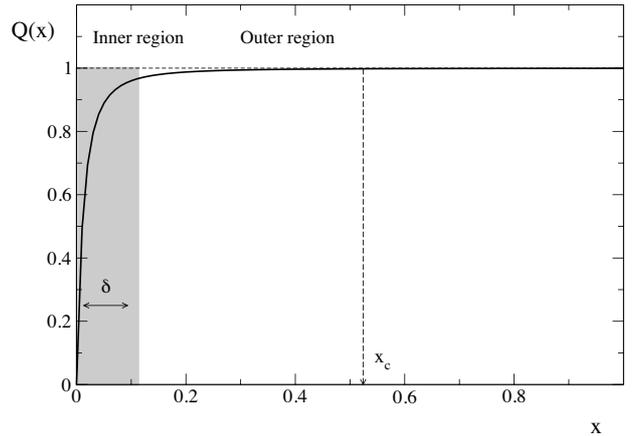}
\caption{Shape of the order parameter function $Q(x)$ for $T\ll 1$
in the Parisi parameterization. The shaded area shows 
the extent of the boundary layer of thickness $\delta\sim T$ as $T\to 0$.}
\label{fig:qx}
\end{figure}

It turns out \cite{Som85,CriRiz02} that as the temperature is decreased towards 
$T=0$ the probability of finding overlaps $Q^{ab}$ sensibly smaller 
than $q_c(T)=O(1)$ vanishes with $T$, while there is a finite probability
$x_c \simeq 0.524\ldots$ that $Q^{ab} \leq q_c(T)$. 
Thus, since $Q(0)=0$, for $T\ll 1$ 
the order parameter function $Q(x)$ in the Parisi parameterization develops a 
{\sl boundary layer} of thickness $\delta \sim T$ close to $x=0$, 
as shown in Fig. \ref{fig:qx}.
From the Figure we see that for very small $T$ the function $Q(x)$ is slowly 
varying for  $\delta \ll x\leq x_c$. However, in the boundary layer 
$0 < x \leq \delta$,
it undergoes an abrupt and rapid change. 
In the limit $T\to 0$ the thickness $\delta\sim T\to 0$ and the order parameter 
function becomes discontinuous at $x=0$.

This behavior of $Q(x)$ for $T\ll 1$ has strong consequences since 
other relevant quantities, such as, e.g, the four-spin correlation entering 
into the Hessian matrix, can be computed
from partial differential equations of the form, 
\begin{equation}
\label{eq:F}
 \dot{F}(x,y) = -\frac{\dot{Q}(x)}{2}
                       \Bigl[
                F''(x,y) + 2\beta\,x\, m(x,y)\,F'(x,y)
                                        \Bigr],
\end{equation}
that gives the function $F(x,y)$ for $x < x^*$ once $F(x^*,y)$ is known.
Details will be given elsewhere \cite{CriDeDomU}.
As usual the ``dot'' and the ``prime'' denote derivative with respect to 
$x$ and  $y$, respectively.
The function $F(x,y)$ is a generic quantity
at the scale $x$ in presence of the frozen field $y$.
The function $m(x,y)$ is the local magnetization and it is itself solution of 
eq. (\ref{eq:F}) with $F(x,y) = m(x,y)$ and initial condition 
$m(1,y) = \tanh \beta y$ \cite{SomDup84}.
In the limit $T\to 0$ we then face a {\sl boundary layer} problem.

Uniform approximate solutions valid for $T\ll 1$ can be constructed 
by using the boundary layer theory, that is by studying the problem 
separately inside ({\sl inner region}) and outside ({\sl outer region}) the
boundary layer \cite{BenOrs99}.
One then introduces the 
notion of the {\sl inner} and {\sl outer} limit of the solution.
The {\sl outer limit} is obtained by choosing a fixed $x$ outside 
the boundary layer,
that is in $\delta \ll x \leq 1$, and allowing $T\to 0$. Similarly the
{\sl inner limit} is obtained by taking $T\to 0$ with $x\leq \delta$. 
This limit
is conveniently expressed introducing an inner variable $a$, 
such as $a=x/\delta$, in terms of which the solution is slowly varying
inside the boundary layer as $T\to 0$.
The inner and outer solutions are then combined together by matching them 
in the {\sl intermediate limit}  $x\to 0$, $x/\delta\to\infty$ and $T\to 0$.

The {\sl inner solution} 
of  $Q(x)$ as $T\to 0$ was first computed by Sommers and 
Dupont in their pioneering work \cite{SomDup84} by
using the inner variable $a$ defined as 
$1 / (dQ(a)/da) = x/T$ and $T\to 0$, 
the so called {\sl Sommers-Dupont gauge}. 
Recently the {\sl inner solution} for the order parameter function 
$Q(x)$ as $T\to 0$ was extensively studied by Oppermann, Sherrington 
and Schmidt \cite{OppShe05,SchOpp08,OppSch08} by using as inner variable
$a = x/T$, as suggested by the Parisi-Toulouse ansatz \cite{ParTho80}.
In both cases one finds that the {\sl inner solution} $Q(a)$ for $T\to 0$ is a 
smooth function of $a$ varying between $0$ and $q_c\simeq 1$.

The {\sl outer solution} was studied by Pankov \cite{Pan06}, who found that
in the outer region one has
\begin{equation}
  Q(x) \sim 1 - c(\beta x)^{-2}, \qquad 
  T\ll x\leq x_c\ \mbox{\rm and } T\to 0
\end{equation}
where $c = 0.4108\ldots$.
The breakpoint $x_c$ is $T$-dependent, however its dependence is
rather weak and $x_c(T) \sim x_c(0) = 0.524\ldots$ is 
a rather good approximation for $T\ll 1$ \cite{CriRiz02}.
From this expression one sees that the variation of the $Q(x)$ in the 
outer region 
$[Q(x_c)-Q(x)]/Q(x) \sim c\left(T/x\right)^2$
is indeed rather weak.

More interestingly in his work Pankov has found that for 
$T\ll x \leq x_c$ and $T\to 0$ both the local magnetization $m(x,y)$
and the distribution function $P(x,y)$ of the frozen field $y$ at scale $x$
loose their explicit dependence on the scale variable $x$. 
This result can be extended to the solution of the generic partial 
differential equation (\ref{eq:F}). By using the same notation as Pankov this
means that in the outer region the solution of eq. (\ref{eq:F}) is
of the form 
\begin{equation}
F(x,y) = \widetilde{F}(z), \ z=\beta x y.
\end{equation}
called {\sl scaling solution} by Pankov.
This {\sl insensitivity} with respect to the scale will allow for 
a complete diagonalization of the Hessian matrix in the outer region.

{\sl The Hessian Matrix.}--With $4$ replicas the Hessian (\ref{eq:hes})
is characterized by $3$ overlaps.
We can distinguish two cases. 
The {\sl Longitudinal-Anomalous} (LA) geometry characterized by
$a\cap b =r$, $c\cap d = s$ and, 
if $r\not= s$, the single cross overlap 
$t = \max [ a\cap c, a\cap d, b\cap c, b\cap d]$:
\begin{equation}
\label{eq:e10}
 M^{ab;cd} = M_t^{r;s}, \qquad r,s,t = 0,1,\ldots R.
\end{equation}
Note that $t=R+1$ if $a=c$ or $a=d$ or $b=c$ or $b=d$.
The {\sl Replicon} geometry where $a\cap b = c\cap d = r$, and
one has the two cross-overlaps
$u = \max[a\cap c, a\cap d]$ and
$v = \max[b\cap c, b\cap d]$
with $u, v\geq r+1$:
\begin{equation}
\label{eq:e11}
 M^{ab;cd} = M_{u;v}^{r;r}, \qquad u,v \geq r+1.
\end{equation}
The Hessian is a $\frac{n(n-1)}{2}\times\frac{n(n-1)}{2}$ symmetric matrix 
that after block-diagonalization 
becomes a string of
$(R+1)\times(R+1)$ blocks along the diagonal for the LA Sector, followed by 
$1\times 1$ fully diagonalized blocks, for the Replicon 
Sector \cite{DeDomCarTem97,TemDeDomKon94,DeDomKonTem98b}.

{\sl Replicon Sector.}--The diagonal elements in the Replicon Sector are 
given by
the {\sl double} Replica Fourier Transform (RFT) of $M_{u;v}^{r;r}$ 
with respect the cross-overlaps $u,v$ \cite{DeDomCarTem97}
\begin{eqnarray}
\label{eq:e13}
 M_{\hat{k},\hat{l}}^{r;r} &=& \sum_{u=k}^{R+1}\sum_{v=l}^{R+1} p_u p_v
         \Bigl[
       M_{u;v}^{r;r} - M_{u-1;v}^{r;r} 
\nonumber\\
&\phantom{=}&\phantom{=====}
- M_{u;v-1}^{r;r} + M_{u-1;v-1}^{r;r}
         \Bigr].
\end{eqnarray}
In the limit $R\to\infty$ the sums
are replaced by integrals and $p_r = x(Q_r)$.
To evaluate $M_{\hat{k};\hat{l}}^{r;r}$ 
we have to compute the matrix elements $M_{u;v}^{r;r}$, that is the four-spin 
average $\langle s^a\,s^b\,s^c\,s^d\rangle$ for the Replicon geometry,
by solving equations of the form (\ref{eq:F}). 
If $r$ lies in the outer region, that is
$p_r = x(Q_r) \gg T$ as $T\to 0$ or, equivalently, for fixed $r\not= 0$ 
and $T\to 0$, then
{\sl insensitivity} with respect to the scale
variable implies that 
$M_{\hat{k};\hat{l}}^{r;r}$ is independent of $k$ and $l$.
Thus by exploiting this {\sl insensitivity} we conclude that
\begin{equation}
M_{\hat{k};\hat{l}}^{r;r} =  M_{\widehat{r+1};\widehat{r+1}}^{r;r}
                          = O\left(\frac{1}{R^2}\right) 
                  \stackrel{R\to\infty}{=} 0.
\end{equation}
The second equality follows from a Ward-Takahashi identity \cite{DeDomKonTem98}.
In the inner region the Replicon spectrum maintains its complexity. However
its relevance becomes less and less important as $T$ approaches zero,
and vanishes in the limit $T\to 0$ when the thickness of the boundary 
shrinks to zero. 
The Replicon spectrum, similarly to the order parameter function $Q(x)$, 
becomes then discontinuous at $x=0$.

{\sl Longitudinal-Anomalous Sector.}--The LA Sector corresponds to 
the $(R+1)\times(R+1)$  diagonal blocks along the diagonal.
Labeling each block with an index $k=0,\ldots, R+1$, the matrix element 
in each block reads \cite{DeDomCarTem97}:
\begin{equation}
\label{eq:LAs}
 \leftidx{_{\rm LA}}M_{\hat{k}}^{r;s} = 
     \Lambda_{\hat{k}}(r)\,\delta_{r,s}^{\rm Kr}
     + \frac{1}{4} M_{\hat{k}}^{r;s}\,\delta_s^{(k)}, 
\quad r,s = 0, \ldots, R
\end{equation}
where $\Lambda_{\hat{k}}(r)$ is a shorthand for
\begin{equation}
  \Lambda_{\hat{k}}(r) = \left\{\begin{array}{ll}
     M_{\hat{k};\widehat{r+1}}^{r;r} & k > r+1, \\
     M_{\widehat{r+1};\widehat{r+1}}^{r;r} & k \leq r+1, \\
\end{array}\right.
\end{equation}
and $\delta_{s}^{(k)} = p_s^{(k)} - p_{s+1}^{(k)}$, $k=0,1,\ldots,R+1$, 
with
\begin{equation}
\label{eq:e35}
p_s^{(k)} = \left\{\begin{array}{ll}
  p_s & s\leq k \\
2 p_s & s > k.
\end{array}\right.
\end{equation}
$M_{\hat{k}}^{r;s}$ is the RFT of the matrix element 
$M_{t}^{r;s}$ with respect the cross-overlap $t$, that is,
\begin{equation}
\label{eq:e29}
 M_{\hat{k}}^{r;s} = \sum_{t=k}^{R+1} p_t^{(r,s)}
               \left(M_t^{r;s} - M_{t-1}^{r;s}\right)
\end{equation}
with , if $r<s$,  
\begin{equation}
p_t^{(r,s)} = \left\{\begin{array}{ll}
  p_t & t\leq r \\
2 p_t & r < t \leq s \\
4 p_t & r < s < t
\end{array}\right.
\end{equation}
If the scale $k$ lies in the outer region then the RFT
$M_{\hat{k}}^{r;s}$ and $\Lambda_{\hat{k}}(r)$ become
insensitive to the value of $k$, and the corresponding blocks
are diagonalized through the eigenvalue
equation\footnote{    
  The boundary term $t=0$ in the RFT is proportional to $p_0 =n$ 
    and vanishes as $n\to 0$. The next
    term is proportional to $p_1\,M_{1}^{r;s}$ since 
    $M_{0}^{r;s} = 0$, and vanishes for $R\to\infty$.
}%
\begin{equation}
\label{eq:e28}
\lambda_{\rm LA} f^r = M_{\widehat{R+1};\widehat{r+1}}^{r;r} f^r 
             +\frac{1}{4}\sum_{s=0}^{R} M_{\widehat{R+1}}^{r;s} \delta_s f^s
\end{equation}
where $\delta_s = p_s - p_{s+1}$.
In the outer region the eigenvectors satisfy 
$f^r \not=0$ if $T\ll x(Q_r) \leq x_c$ as $T\to 0$, and zero otherwise.
Thus the eigenvalue equation becomes
\begin{equation}
\label{eq:e30}
 \lambda_{\rm LA} f^r = \frac{1}{4} M_{\widehat{R+1}}^{R;R} 
 \sum_{s=\overline{r}}^{R} \delta_s f^s,
\qquad r = \overline{r},\ldots, R.
\end{equation}
where $\overline{r}$ is the lower bound of the outer region, that is
$x(Q_{\overline{r}}) = \overline{x} \sim \delta$
as $T\to 0$.
The diagonal Replicon contribution vanishes for
$R\to\infty$, as ensured by the Ward-Takahashi 
identity, and does not contribute. This equation has two distinct solutions. 
The first
\begin{equation}
\label{eq:la0}
\lambda_{\rm LA} = 0
\end{equation} 
for $\sum_{s=\overline{r}}^{R}\delta_s f^s=0$
and
\begin{eqnarray}
\label{eq:la1}
\lambda_{\rm LA}&=& \frac{1}{4}\left(\sum_{s=\overline{r}}^{R}\delta_s\right)\, 
            M_{\widehat{R+1}}^{R;R} 
\nonumber\\
            &=&
            (\overline{x} - 1) 
            \bigl(1 - \beta^2\,(1-q_c(T)\bigr)
\nonumber\\
            &=&
            (\alpha - 1) + O(T), 
\quad T\to 0
\end{eqnarray}
for $\sum_{s=\overline{r}}^{R}\delta_s f^s\not=0$.
The last equality follows from $q_c(T) = 1 - \alpha\, T^2 + {\cal O}(T^3)$ 
as $T\to 0$ with
$\alpha = 1.575\ldots$ \cite{CriRiz02}.
In the inner region, where the LA spectrum maintains the RSB structure,
 the solutions are smooth functions of the inner
variable even for $T\to 0$, while the width of the boundary layer vanishes 
in this limit. Therefore for $T\to 0$
the eigenvalues (\ref{eq:la0}) and (\ref{eq:la1}) cover the whole
LA spectrum, with a discontinuity at $x=0$.

{\sl Conclusions.}--To summarize, we have presented the analysis
of the spectrum of the Hessian for the Parisi solution of the SK model in 
the limit $T\ll 1$. It has been long known that in this regime two distinct 
regions can be identified according to the variation of the
order parameter function $Q(x)$ with $x$. 
The structure of the spectrum of the Hessian was, however, 
never studied.
In this Note we have shown that the behavior of $Q(x)$ for $T\ll 1$ 
has strong consequences on the eigenvalue spectrum.
In the first region $x \leq \delta \sim T$, where $Q(x)$ varies rapidly 
from $Q(0)=0$ up to $Q(x) \sim q_c(T)\sim 1$, the spectrum maintains the 
complex structure found close to the critical temperature $T_c$ 
for the full RSB state.
We then call this region the 
{\sl RSB-like} regime.
In the second region, $T \ll x \leq x_c$ with $x_c\sim 0.575\ldots$
where $Q(x)$ is slowly varying, however, the eigenvalue spectrum has a 
completely different aspect. 
The bands observed in the RSB regime
collapse and only two distinct eigenvalues are found: a null one and
a positive one. This ensures that 
the Parisi solution of the SK model remains stable down to zero temperature.
Massless propagators arise from Replicon geometry, with 
Ward-Takahashi identities protecting masslessness. Note, however, that 
the zero modes arise also from LA geometry, that is without protection of 
the Ward-Takahashi identities.

We observe that for $T\ll 1$ the order parameter function is almost 
constant for $T\ll x\leq x_c$, the variation being indeed of order 
$[Q(x_c)-Q(x)]/Q(x) = {\cal O}\left((T/x)^2\right)$. Thus in this region
we have a marginally stable (almost) replica symmetric solution, that becomes 
a genuine replica symmetric solution in the limit $T\to 0$, with self-averaging 
trivially restored.
It is worth to remind 
that the stability analysis of the replica symmetric solution
also leads to two eigenvalues, one of which is zero to the lowest order
in $T_c-T$ (and negative to higher order), and the other positive.

Recently Aspelmeier, Moore and Young \cite{AspMooYou03}
have found that the interface free energy associated with the change from 
periodic to antiperiodic boundary conditions in finite dimensional 
spin glass does not follow the scaling form $L^\theta f(L/M)$, typical of a 
droplet scenario,
if the state is described by a RSB scenario.
Here $\theta$ is the stiffness exponent, $L$ the length of the system along
which the periodic/antiperiodic boundary conditions are applied, and $M$ 
the length in the perpendicular directions. However the scaling form 
is obeyed 
if the state is described 
by a marginally stable replica symmetric solution. 
These results were found using the truncated model, an approximation of the SK 
model valid close to $T_c$ and used here to work with explicit solutions. The
main conclusion should nevertheless be also valid for the full SK 
model \cite{AspMooYou03},
implying that, in the region
$T\ll x \leq x_c$ the Replica Symmetric description prevails and the SK model 
is in a {\sl droplet-like} regime.

Concerning the multiplicity of the eigenvalues we observe that 
in each Sector, Replicon and LA, one has to separate the contribution from 
the RSB-like and the droplet-like regions. The former is proportional to
the width $\delta$ of the region. Therefore in the limit $T\to 0$ the
contribution from the RSB-like region vanishes, and one has the usual
Replicon and LA multiplicities for the droplet-like region.

Since 
in the limit $T\to 0$ the domain of the RSB-like regime 
shrinks to zero, and only the droplet-like regime survives,
it can be viewed as a cross-over between the two scenarios.
It is interesting to note that in the dynamical approach small
values of $x$ correspond to large time scales. As a consequence 
this implies that for $T\ll 1$  the RSB scenario is seen 
on very very long time scales, while on shorter time scales a more 
droplet-like scenario is observed. 

We conclude by noticing that while these results strongly 
suggest a cross-over 
between RSB and droplet descriptions
in spin glasses, to have a better understanding of the behavior
of finite dimensional systems loop corrections to the mean-field 
propagators must be considered \cite{BraMoo86},
a task beyond the scope of this Note.

\acknowledgments
The authors acknowledge useful discussions with M. A. Moore, 
R. Oppermann, T. Sarlat and A. P. Young.
A.C. acknowledges hospitality and support from 
IPhT of CEA, where part of this work was done.

\end{document}